\documentclass[amsmath,amssymb,aps,preprint]{revtex4-1}

\usepackage{graphicx}
\usepackage{dcolumn}
\usepackage{bm}

\usepackage{ulem}
\usepackage{color}
\usepackage{xcolor}

\begin{document}


\title{Linking Langevin equation to scaling properties of space plasma turbulence at sub-ion scales}


\author{Simone Benella$^1$, Mirko Stumpo$^1$, Tommaso Alberti$^2$, Oreste Pezzi$^{3,1}$, Emanuele Papini$^1$, Emiliya Yordanova$^4$, Francesco Valentini$^5$ and Giuseppe Consolini$^1$}

\affiliation{$^1$Istituto di Astrofisica e Planetologia Spaziali, Istituto Nazionale di Astrofisica, via del Fosso del Cavaliere 100, I-00133, Rome, Italy\\
$^2$Istituto Nazionale di Geofisica e Vulcanologia, Via di Vigna Murata, 605, I-00143, Rome, Italy\\$^3$Istituto per la Scienza e Tecnologia dei Plasmi, Consiglio Nazionale delle Ricerche, Via Amendola 122/D, I-70126 Bari, Italy\\$^4$Swedish Institute for Space Physics, Uppsala, Sweden\\$^5$Dipartimento di Fisica, Università della Calabria, Ponte P. Bucci, Cubo 31C, I-87036 Arcavacata di Rende (CS), Italy}
\date{\today}

\begin{abstract}
    Current understanding of the kinetic-scale turbulence in weakly-collisional plasmas still remains elusive. We employ a general framework in which the turbulent energy transfer is envisioned as a scale-to-scale Langevin process. Fluctuations in the sub-ion range show a global scale invariance, thus suggesting a homogeneous energy repartition. In this Letter, we interpret such a feature by linking the drift term of the Langevin equation to scaling properties of fluctuations. Theoretical expectations are verified on solar wind observations and numerical simulations thus giving relevance to the proposed framework for understanding kinetic-scale turbulence in space plasmas.
\end{abstract}

\maketitle



Space and astrophysical plasmas, such as the solar and stellar winds, the planetary magnetospheres, and the interstellar medium, often exhibit a strongly turbulent dynamics. The energy of fluctuations cross-scale transfers over a vast range of spatial and temporal scales and diverse structures, including current sheets, magnetic islands, and vortices, emerge. Since about sixty years, a large fleet of spacecraft has been providing in-situ measurements of the solar wind plasma and the interplanetary magnetic field \cite{bruno16,chasapis2017}. These observations have allowed us to investigate in great detail turbulence in space plasmas, whose study is of pivotal importance also for clarifying the dynamics of far astrophysical objects not accessible by in-situ spacecraft.

Owing to their weak collisionality, understanding the properties of turbulence at scales similar or smaller than the inertial ion scales, henceforth referred as sub-ion scales, is essential for clarifying the energy transfer and dissipation mechanisms in space and astrophysical plasmas.
Indeed, at sub-ion scales, the solar wind dynamics is a complex puzzle composed of several intertwined plasma processes. Such processes generally manifest in the entire phase space, since inter-particle collisions are infrequent and plasmas can be far from the thermal equilibrium \citep{cassak2023}.

Historically, the investigation of plasma dynamics at sub-ion scales has aimed at identifying the plasma processes successfully explaining in-situ observations. These include kinetic-scale fluctuations (e.g., kinetic-Alfvén and whistler waves) \cite{Alexandrova2008,sahraoui2009evicence,chen2010,chen2013}, microinstabilities \cite{matteini2012}, magnetic reconnection \cite{zenitani2011,wan2015}, energy transfer highlighted by pressure-strain interactions \cite{Bandyopadhyay2020}, Landau damping \cite{chen2019} etc., \cite[see][for a review]{Verscharen2019}. These mechanisms often depend on different plasma parameters (e.g., the plasma $\beta$) and, hence, their applicability is specialized to specific plasma conditions. Moreover, some of these mechanisms, such as kinetic-scale fluctuations, microinstabilities, and Landau damping, are often introduced within a linear or quasi-linear framework, thus possibly making their application less suitable in cases of strong turbulence, as observed in the solar wind \cite{matthaeus2014}.

A complementary and additional perspective proposes to interpret space plasma turbulence through a small set of macroscopic variables, whose fluctuations are not negligible but rather control the system dynamics \cite{sekimoto2010}.
Such a definition sets the ground for a general and interdisciplinary framework whose underlying idea is that an observed coarse-grained dynamics can be modeled as a stochastic processes. A stunning example of this approach is represented by hydrodynamic turbulence, where longitudinal velocity fluctuations at different scales can be described by means of a stochastic approach \cite[e.g.,][]{Faranda17,Dubrulle19,Alberti23}, mostly based on the generalized Langevin equation \cite{friedrich1997,davoudi1999,renner2001evidence,milan2013,Nickelsen2013,fuchs2022}.
From this point of view, the scale-by-scale evolution of the probability distribution functions (PDFs) of the longitudinal increments is governed by the Fokker-Planck equation (FPE). A similar description has been also applied to turbulence in space plasmas. An earlier work by \citet{strumik2008testing} showed that the inertial-range cascade of the solar wind can be properly framed in this scenario. A step forward has been made by proving that this formalism applies to sub-ion scales, thanks to unprecedented high-resolution magnetic field observations 
\cite{benella2022markovian, benella2022kramers,macek2022mms}. These studies confirm that the FPE is accurate in reproducing the observed PDFs of magnetic field fluctuations, 
and show that PDFs outline a globally self-similar scenario \cite{kiyani2009global,osman2015,benella2022markovian,benella2022kramers,macek2022mms}.

A common way to investigate solar wind turbulence is to analyze high-order statistics of magnetic field fluctuations, viz., the structure functions $S^{(q)}_r=\langle b_r^q\rangle$, where
\begin{equation}
    b_r \doteq B(x+r) - B(x)
    \label{eq:increments}
\end{equation}
are the increments of the magnetic field $B(x)$ at the spatial scale $r$. 
Fluctuations $b_r$ are \textit{globally} scale invariant if
\begin{equation}
    r \to \lambda r \Rightarrow b_{\lambda r}=\lambda^{h} b_r,
    \label{eq:stochproc}
\end{equation}
where $h$ is a scaling exponent and $\lambda \in \mathbb{R}^+$ is a dilatation/contraction parameter representing a zoom-in/out procedure through magnetic field structures. The basic idea is that each structure at a scale decays into smaller-scale structures in a scale-invariant way. Indeed, Eq. (\ref{eq:stochproc}) implies that structure functions exhibit a power-law trend as a function of $r$, i.e., $S^{(q)}_r\sim r^{\zeta_q}$, with the scaling exponents $\zeta_q$. The space-time distribution of turbulent structures reflects the properties of energy dissipation rate and related underlying processes \cite{frisch1995}. 
However, there are no fully deductive theories that are able to derive the scaling laws, viz., the behavior of the scaling exponents, starting from dynamic equations. For this reason, phenomenological theories, which introduce additional hypothesis based on experimental evidences, are widely used. In this context, the method described in this Letter ranks as a novel phenomenological perspective of sub-ion scale plasma turbulence.

The transformation introduced in Eq. (\ref{eq:stochproc}) is satisfied by $b_r\sim r^{h}$ \cite{frisch1995}. In the multifractal theory of magnetohydrodynamic (MHD) turbulence, $h$ is the generalized H\"older exponent $h(x,r)$ \cite{holder} and thus magnetic field fluctuations are \textit{locally} scale-invariant, $b_r \sim r^{h(x,r)}$.
Here, $h(x,r)$ represents a fluctuating quantity \cite{landau2013fluid}, hence the global scale-invariance holds only on average \cite{frisch1995} and can be expressed in terms of the most probable H\"older exponent $h_0$. For a globally scale-invariant dynamics, a relation between $h_0$ and the spectral slope $\beta$ of the Fourier power spectral density can be derived
\begin{equation}
    \beta = 2 h_0 + 1,
    \label{eq:spec}
\end{equation}
being $\beta \in (1, 3)$ for the non-stationary process $B(x)$ with stationary increments $b_r$ \cite{flandrin1989spectrum}.

In terms of FPE, the scale-evolution equation of the $q$-th order structure function can be written as \cite{benzi1993,renner2001experimental,arneodo2008}
\begin{equation}
    -\frac{\partial}{\partial r}S^{(q)}_r= q\langle b_r^{q-1}D^{(1)}(b_r)\rangle+ q(q-1)\langle  b_r^{q-2}D^{(2)}(b_r)\rangle.
    \label{eq:sq_fp}
\end{equation}
This scale-evolution equation depends on the two Kramers-Moyal (KM) coefficients which define the stochastic process. For the solar wind, the following parameterization of the KM coefficients correctly reproduce the fluctuation statistics at sub-ion scales \cite{benella2022markovian,benella2022kramers,macek2022mms}:
\begin{equation}
    \begin{aligned}
        D^{(1)}(b_r)=-\gamma(r)\, b_r \\
        D^{(2)}(b_r)=\alpha(r) + \delta(r)\, b_r^2.
    \end{aligned}
    \label{eq:d12}
\end{equation}
By inserting Eqs. (\ref{eq:d12}) in Eq. (\ref{eq:sq_fp}), a direct relation between KM coefficient parameters and the local scaling exponent $\xi_q(r)$ can be drawn \cite{renner2001experimental,reinke2018on}:
\begin{equation}
    \xi_q(r)=\frac{r}{S^{(q)}_r}\frac{\partial S^{(q)}_r}{\partial r}=r q\biggl( \gamma(r)+(1-q) \biggl( \delta(r)+\frac{S^{(q-2)}_r}{S^{(q)}_r}\alpha(r) \biggr)\biggr).
    \label{eq:xiq}
\end{equation}
In the framework of our model, it means that the evolution in scale of the fluctuating trajectories $b_r$ can be described according with the Langevin equation \cite{risken1996fokker}:
\begin{equation}
    -\frac{\partial b_r}{\partial r} = D^{(1)}(b_r,r)+\sqrt{2D^{(2)}(b_r,r)}\eta(r),
    \label{eq:lang}
\end{equation}
where $\eta(r)$ is a unit-variance Gaussian noise term such that $\langle \eta(r)\eta(r')\rangle=\delta(r-r')$. Eq. (\ref{eq:lang}) indicates that the dynamics of magnetic field increments as a function of the scale $r$ is the result of the superposition of two mechanisms: the \textit{drift} which models the self-similar decay of magnetic field structures from larger to smaller ones and the \textit{diffusion} that accounts for the stochastic repartition of energy during the structure breakdown process.

This scenario applies in the general framework of turbulence. In the following we specialize our discussion to the sub-ion scale dynamics by considering three different data samples: super-Alfvénic solar wind observations gathered by PSP during its first perihelion 
\cite{chhiber2021subproton}, a high-speed stream observed by the ESA/Cluster mission in the near-Earth solar wind 
\cite{alberti2019multifractal}, and a data sample obtained from Eulerian Hybrid Vlasov-Maxwell (HVM) simulations of decaying plasma turbulence 
\cite{pezzi2019proton,pezzi2021dissipation}. The HVM numerical framework has been widely adopted to describe significant features of turbulence in space plasmas \citep{servidio2015kinetic,cerri2017kinetic,franci2018solarwind}, including the transition to a monofractal scaling when approaching sub-ion scales \citep{leonardis2016}.
\begin{figure}
    \centering
    \includegraphics[width=0.6\columnwidth]{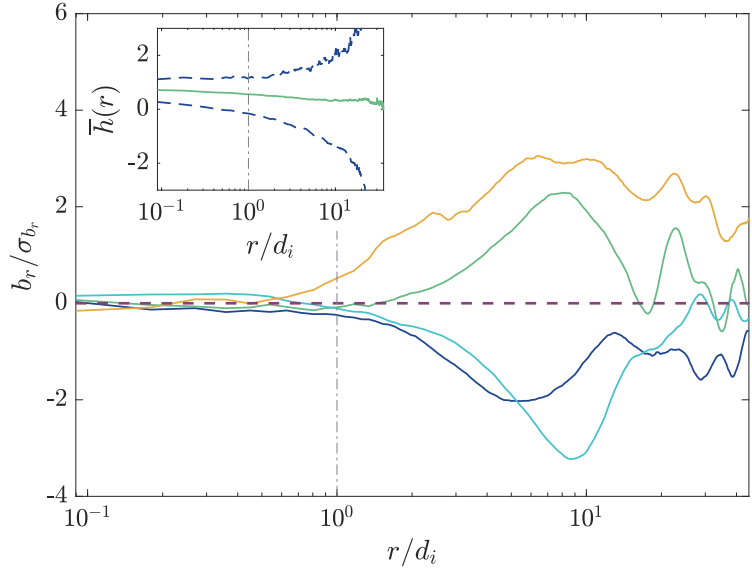}
    \caption{Fluctuating magnetic field trajectories normalized to the standard deviation scale-by-scale from the PSP data sample (color lines). The inset shows the averaged H\"older exponent $\overline{h}(r)$ as a function of $r$ (solid line) along with 95\% confidence bounds (dashed lines) for the PSP data sample. Vertical lines indicate the ion inertial length.}
    \label{fig:1}
\end{figure}
The data selected for the analysis show a similar level of fluctuation $b_\text{rms}/B_0\sim0.5$, being $B_0$ the mean magnetic field, and different values of the proton plasma beta (0.4 for PSP, 1.5 for Cluster and 2 for HVM).
The analysis of spacecraft measurements is carried out by assuming the Taylor's hypothesis: $r=\langle V\rangle\tau$, where $\tau$ indicates the time scale and $\langle V\rangle$ the mean solar wind speed \cite{narita2013,chhiber2019}.
%
Plasma turbulence in the solar wind is known to be strongly anisotropic and the energy cascade develops mostly in the transverse directions with respect to the mean local magnetic field \cite{shebalin1983anisotropy,matthaeus1996anisotropic,verma2004statistical, matthaeus2012local}. Hence, we analyze fluctuations along one of the transverse magnetic field directions for each data sample considered here. For \textit{in-situ} observations we compute increments along the maximum variance component $b_r= b_{r,\max}$, while in the case of HVM simulations we consider increments along the $x$-component $b_r = b_{r,x}$ since the mean field is directed along the $z$-direction \cite{pezzi2019proton}. 

The first step in our reasoning consists in assessing the different role played by the deterministic and the stochastic terms in Eq. (\ref{eq:lang}) at different scales.
To this purpose we define a set of \textit{trajectories} $b_r$, see Eq. (\ref{eq:increments}), for different starting points $x$.
For a nearly constant value of the H\"older exponent, say $h_0$, the amplitude of fluctuations is expected to decrease as a power law in $r$. If we normalize each trajectory by the standard deviation $\sigma_{b_r}$, which then scales as $r^{h_0}$, we expect to observe constant and scale-independent values of the normalized fluctuations. In this view, Figure \ref{fig:1} reports a few normalized trajectories which allow us to identify two different dynamical regimes: i) the sub-ion range where normalized fluctuations show a nearly constant value; ii) the lower end of the inertial range, e.g., at scales $\gtrsim10d_i$, where trajectories exhibit stochastic fluctuations. The different level of stochasticity of the magnetic field fluctuations at different scales can be quantified by estimating the local H\"older exponent. The inset of Figure \ref{fig:1} reports the scale-by-scale averaged H\"older exponent $\overline{h}(r)$ (solid line) along with the 95\% confidence interval bounds (dashed lines). The stochastic character of turbulent fluctuations is related to the spreading of the local H\"older exponent (i.e., spreading of the confidence bounds) when approaching the inertial range ($r > 10 d_i$). 
In the context of Langevin equation, this scenario suggests that the diffusion term does not significantly contribute to the dynamics at sub-ion scales, whereas it plays an important role in the inertial range. Results of Figure \ref{fig:1} are derived from PSP data but, analogous considerations can be also extended to the other data samples (not shown). Based on this observation, we introduce the non-diffusive limit of the Langevin equation in order to model sub-ion scale fluctuations.

Hence, we assume that the small-scale dynamics is governed only by the first-order KM coefficient, viz., $D^{(1)}\gg D^{(2)}$, which in terms of the parametrization introduced in Eqs. (\ref{eq:d12}) corresponds to neglecting the parameters $\alpha(r)$ and $\delta(r)$. Furthermore, $\gamma(r)$ follows a power-law behavior at sub-ion scales \cite{benella2022markovian,benella2022kramers,macek2022mms}, i.e.,
\begin{equation}
    \gamma(r)=h_0r^\mu,
    \label{eq:gamma_par}
\end{equation}
allowing us to derive the coefficients $h_0$ and the exponents $\mu$ directly from data. Their values are listed in Table \ref{tab:1}.
In the non-diffusive limit, Eq. (\ref{eq:xiq}) becomes
\begin{equation}
    \xi_q (r)\sim r q\,\gamma(r)=qh_0r^{1+\mu}.
    \label{eq:k41}
\end{equation}
For both solar wind data and numerical simulations $\mu\sim-1$ (see Table \ref{tab:1}), then Eq. (\ref{eq:k41}) reads
\begin{equation}
    \zeta_q \equiv \xi_q(r) = q h_0,
    \label{eq:z_q}
\end{equation}
where $h_0$, in this fashion, corresponds to the most probable H\"older exponent. This means that the condition $D^{(1)}\gg D^{(2)}$ implies a \textit{global scale-invariance}. This represents the central result of this work since it establishes a direct link between the drift term at sub-ion scales and the most probable H\"older exponent of the multifractal theory. Such a connection will be now verified by exploiting both \textit{in-situ} observations and numerical simulations in a two-fold sense: 
on a statistical level, e.g., through structure functions, and on an individual level, e.g., by considering single fluctuating trajectories.

\begin{figure}
    \centering
    \includegraphics[width=0.6\columnwidth]{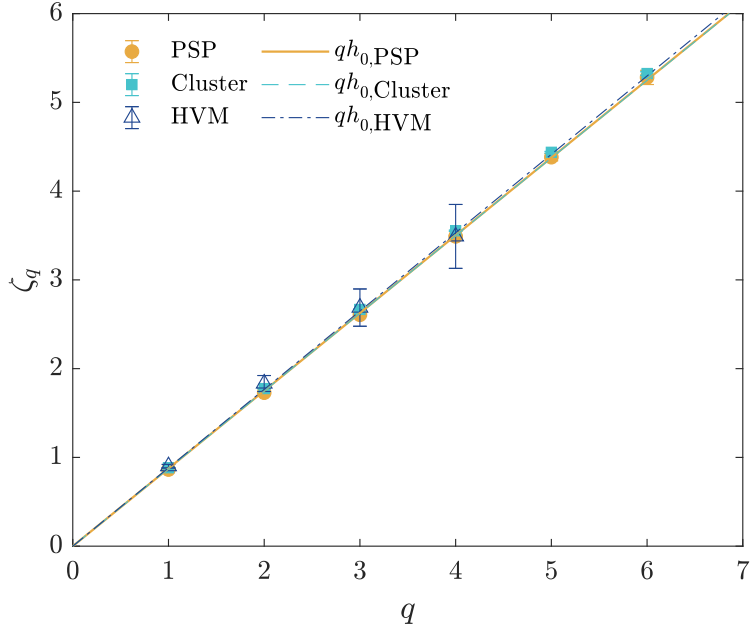}
    \caption{Scaling exponents $\zeta_q$ as a function of the order $q$ of the structure functions (markers). Straight lines represent the estimation provided by Eq. (\ref{eq:z_q}) by using the values of Table \ref{tab:1}.}
    \label{fig:3}
\end{figure}

Scaling exponents can be derived from solar wind observations through the structure function analysis within the sub-ion range.
The scaling exponents obtained from the three data samples are here compared with those obtained through Eq. (\ref{eq:z_q}), Figure \ref{fig:3}. An excellent agreement between observations and model predictions is found, strongly supporting the established link between the first-order KM coefficient and the H\"older exponent $h_0$. Concerning the spectral analysis, we can use Eq. (\ref{eq:spec}) to estimate the spectral slope corresponding to the observed most probable H\"older exponent $h_0$ from the three data samples. By using the values estimated from the drift coefficient, reported in Table \ref{tab:1}, we obtain $\beta\sim[2.7,2.8]$, which is in agreement with typical findings in the sub-ion range \cite{sahraoui2013scaling,carbone2022high}.
\begin{table}[]
    \centering
    \begin{tabular}{lccc}    \toprule
        & PSP & Cluster & HVM \\\hline
        $h_0$   & $0.87\pm0.06$ & $0.87\pm0.03$ & $0.88\pm0.52$\\
        $\mu$ & $-1.02\pm0.01$ & $-1.04\pm0.01$ & $-1.04\pm0.22$\\\hline
    \end{tabular}
    \caption{Fitted $h_0$ and $\mu$ parameters appearing in Eq. (\ref{eq:gamma_par}). Errors represent fit 95\% confidence bounds.}
    \label{tab:1}
\end{table}
\begin{figure*}
    \centering
    \includegraphics[width=\textwidth]{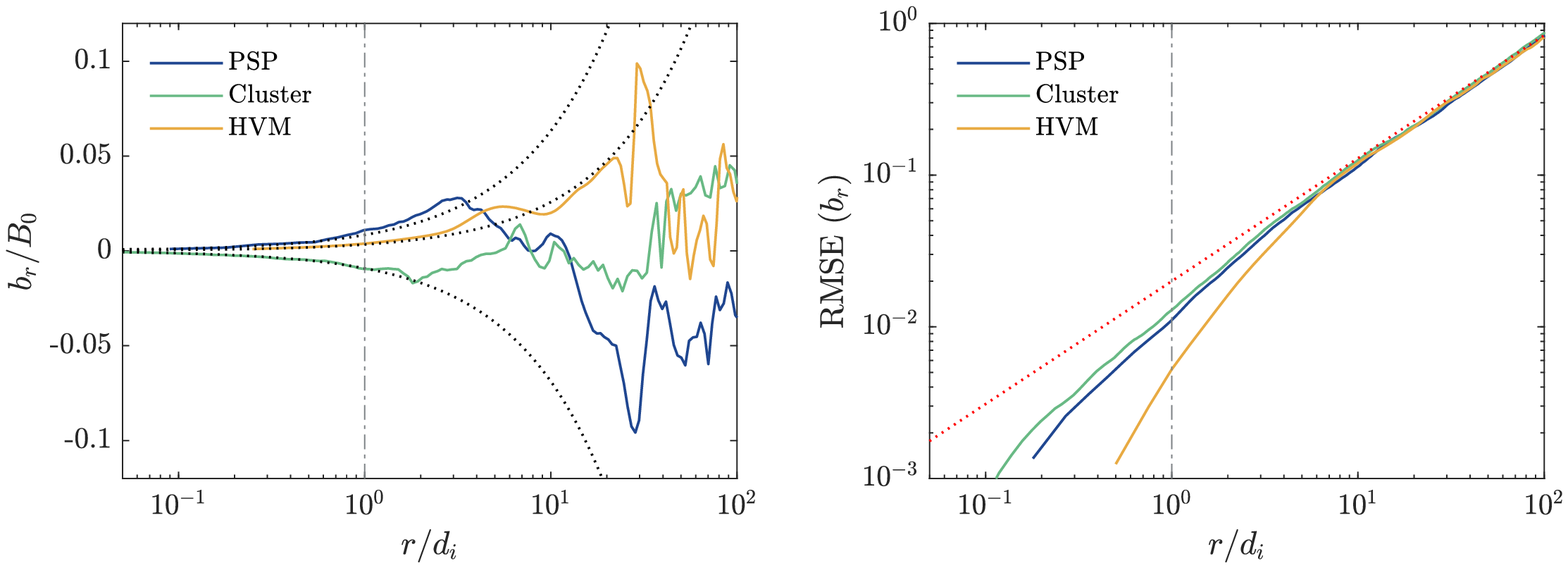}
    \caption{Comparison between three trajectories selected from different data samples (left panel) and the corresponding non-diffusive trajectories $b_r^*$ (black dotted lines). Root mean square error of the fluctuations $b_r$ with respect to $b_r^*$ as a function of the scale (right panel). The red dotted line indicates a reference power-law trend. The scales are reported in units of $d_i$ and the magnetic field is normalized by the mean field value $B_0$ for clarity of presentation.}
    \label{fig:4}
\end{figure*}

The linear trend of the scaling exponents $\zeta_q$ as a function of the order of the structure function $q$ observed from data suggests that magnetic field fluctuations should exhibit a weak stochastic behavior at sub-ion scales. In fact, the non-diffusive limit of Eq. (\ref{eq:lang}) corresponds to a damping equation with a scale dependent drift term
\begin{equation}
    \frac{\partial b_r^*}{\partial r}=\gamma(r)b_r^*=\frac{h_0 b_r^*}{r},
    \label{eq:damp_diff}
\end{equation}
whose solution is:
\begin{equation}
    b_r^* = b_0\biggl(\frac{r}{r_0}\biggr)^{h_0},
    \label{eq:damp}
\end{equation}
where $b_r^*$ indicates the deterministic solution and $b_0$ is the initial condition, i.e., a value of the fluctuation at the scale $r_0$.
The left panel of Figure \ref{fig:4} shows a comparison of the power-law damping predicted by Eq. (\ref{eq:damp}) with a set of fluctuating trajectories from three different data samples. This figure points out that the non-diffusive approximation is quite satisfactory for $r\ll d_i$, whereas the diffusion starts to significantly affect the dynamics of the magnetic field fluctuations at $r\gtrsim d_i$. The root mean square error between the whole set of measured trajectories and the corresponding model predictions is reported in Figure \ref{fig:4} (right panel). This figure shows that errors decrease more rapidly when approaching sub-ion scales for all three data samples.

The proposed approach allows us to unveil connections between the global statistical features of the plasma dynamics and the peculiar characteristics of each realization of the cascading process. For instance, the H\"older exponent $h_0$ quantifies the regularity (i.e., the roughness) of the topology of the fluctuations associated with a dynamical process \cite{nguyen20}: irregular bursts are characterized by $h_0 \to 0$, whereas a field of increasingly regular fluctuations is associated with larger $h_0$ and leans towards an homogeneous fractal structure when $h_0\to1$.
By looking at Figure \ref{fig:4} we can observe that the deterministic-like damping shown by empirical fluctuating trajectories is characterized by a weakly fluctuating trend that corresponds to a well-defined sign of the fluctuation amplitude among different scales. The high values obtained for $h_0\sim[0.87,0.88]$ 
represent a natural consequence of this intuitive framework.

In the case of the Langevin equation, the non-diffusive limit is achieved by simply neglecting the noise term  and then reducing the stochastic equation to a deterministic one. The same procedure cannot be applied when dealing with the FPE, for which the non-diffusive limit is singular \cite{gardiner1985handbook, vankampen1992}. 
In this case 
the correct way to perform this limit is to introduce a new variable $y$ such that $b_r =b^*_r+\epsilon y$, where $\epsilon$ is an arbitrarily small parameter. Hence, the non-diffusive limit corresponds to $\epsilon\to0$ in which case the stochastic fluctuation vanishes and $b_r$ approaches the deterministic solution $b_r^*$.
This approximation represents an ideal limit in which the FPE predicts weak fluctuations around the deterministic damping \cite{vankampen1992}. Such scenario seems to be fairly accurate by inspecting Figure \ref{fig:4}, where the trajectories exhibit weak fluctuations around the asymptotic solution at small scales.

All the evidences reported in this work are consistent with 
a homogeneous and globally self-similar repartition of energy through the scales in the sub-ion range. As a consequence, the typical multifractal character of the non-homogeneuous energy repartition observed in the inertial range, here enclosed in the diffusion term of the Langevin equation, does not represent a major contribution at these scales. The result is an unstructured fluctuation field suggesting that typical space-time inhomogeneities associated with turbulent structures have sizes larger than $d_i$. This is supported by the linear scaling law of $\zeta_q$, here derived in the non-diffusive limit, and by the fact that trajectories at sub-ion scales are captured with a single exponent $h_0$. In this context, our newly introduced framework constitutes a complementary view with respect to the existing phenomenological models, allowing us to recover important results of solar wind turbulence at sub-ion scales and to provide new constraints for future model developments.

\begin{acknowledgements}
    This work is supported by the mini-grant ``The solar wind: a paradigm for complex system dynamics'' financed by the National Institute for Astrophysics under the call ``Fundamental Research 2022''. E.Y. work was funded by the Swedish Space Agency, grant 192/20.
    PSP data used in this study are available at the NASA Space Physics Data Facility (SPDF), \url{https://spdf.gsfc.nasa.gov/index.html}. The authors acknowledge the contributions of the FIELDS team to the Parker Solar Probe mission. The authors acknowledge the Cluster FGM and STAFF P.I.s and teams and the ESA-Cluster Science Archive for making available the data used in this work. The FGM and STAFF data were combined using the IRFU-Matlab analysis package available at \url{https://github.com/irfu/irfu-matlab}.
\end{acknowledgements}


\bibliography{LinkingLangevin}

\end{document}